\documentstyle[twoside,epsfig,amsmath,amssymb]{article}
 
\newcommand{\Tr}{\hbox{Tr}} 
 
\newcommand{\I}{{\rm i}} 
\newcommand{\E}{{\rm e}} 
\newcommand{\Geff}{\Gamma_{\text{eff}}} 
\newcommand{\Leff}{{\cal L}_{\text{eff}}} 
\newcommand{\re}[1]{~(\ref{#1})} 
\newcommand{\fss}[1]{#1\!\!\!/}
\catcode`\@=11
\long\def\@makefntext#1{
\protect\noindent \hbox to 3.2pt {\hskip-.9pt  
$^{{\eightrm\@thefnmark}}$\hfil}#1\hfill}               

\def\thefootnote{\fnsymbol{footnote}}
\def\@makefnmark{\hbox to 0pt{$^{\@thefnmark}$\hss}}    
        
\def\ps@myheadings{\let\@mkboth\@gobbletwo
\def\@oddhead{\hbox{}
\rightmark\hfil\eightrm\thepage}   
\def\@oddfoot{}\def\@evenhead{\eightrm\thepage\hfil
\leftmark\hbox{}}\def\@evenfoot{}
\def\sectionmark##1{}\def\subsectionmark##1{}}

\oddsidemargin=\evensidemargin
\renewcommand{\thefootnote}{\fnsymbol{footnote}}

\newcounter{sectionc}\newcounter{subsectionc}\newcounter{subsubsectionc}
\renewcommand{\section}[1] {\vspace{12pt}\addtocounter{sectionc}{1} 
\setcounter{subsectionc}{0}\setcounter{subsubsectionc}{0}\noindent 
        {\tenbf\thesectionc. #1}\par\vspace{5pt}}
\renewcommand{\subsection}[1] {\vspace{12pt}\addtocounter{subsectionc}{1} 
        \setcounter{subsubsectionc}{0}\noindent 
        {\bf\thesectionc.\thesubsectionc. {\kern1pt \bfit #1}}\par\vspace{5pt}}
\renewcommand{\subsubsection}[1] {\vspace{12pt}\addtocounter{subsubsectionc}{1}
        \noindent{\tenrm\thesectionc.\thesubsectionc.\thesubsubsectionc.
        {\kern1pt \tenit #1}}\par\vspace{5pt}}
\newcommand{\nonumsection}[1] {\vspace{12pt}\noindent{\tenbf #1}
        \par\vspace{5pt}}

\newcounter{appendixc}
\newcounter{subappendixc}[appendixc]
\newcounter{subsubappendixc}[subappendixc]
\renewcommand{\thesubappendixc}{\Alph{appendixc}.\arabic{subappendixc}}
\renewcommand{\thesubsubappendixc}
        {\Alph{appendixc}.\arabic{subappendixc}.\arabic{subsubappendixc}}

\renewcommand{\appendix}[1] {\vspace{12pt}
        \refstepcounter{appendixc}
        \setcounter{figure}{0}
        \setcounter{table}{0}
        \setcounter{lemma}{0}
        \setcounter{theorem}{0}
        \setcounter{corollary}{0}
        \setcounter{definition}{0}
        \setcounter{equation}{0}
        \renewcommand{\thefigure}{\Alph{appendixc}.\arabic{figure}}
        \renewcommand{\thetable}{\Alph{appendixc}.\arabic{table}}
        \renewcommand{\theappendixc}{\Alph{appendixc}}
        \renewcommand{\thelemma}{\Alph{appendixc}.\arabic{lemma}}
        \renewcommand{\thetheorem}{\Alph{appendixc}.\arabic{theorem}}
        \renewcommand{\thedefinition}{\Alph{appendixc}.\arabic{definition}}
        \renewcommand{\thecorollary}{\Alph{appendixc}.\arabic{corollary}}
        \renewcommand{\theequation}{\Alph{appendixc}.\arabic{equation}}
        \noindent{\tenbf Appendix \theappendixc #1}\par\vspace{5pt}}
\newcommand{\subappendix}[1] {\vspace{12pt}
        \refstepcounter{subappendixc}
        \noindent{\bf Appendix \thesubappendixc. {\kern1pt \bfit #1}}
        \par\vspace{5pt}}
\newcommand{\subsubappendix}[1] {\vspace{12pt}
        \refstepcounter{subsubappendixc}
        \noindent{\rm Appendix \thesubsubappendixc. {\kern1pt \tenit #1}}
        \par\vspace{5pt}}

\newcommand{\textlineskip}{\baselineskip=13pt}
\newcommand{\smalllineskip}{\baselineskip=10pt}

\def\eightcirc{
\begin{picture}(0,0)
\put(4.4,1.8){\circle{6.5}}
\end{picture}}
\def\eightcopyright{\eightcirc\kern2.7pt\hbox{\eightrm c}} 

\newcommand{\copyrightheading}[1]
        {\vspace*{-2.5cm}\smalllineskip{\flushleft
        {\footnotesize 
        \hfill CERN-TH-2001-368}\\
         }}

\def\abstracts#1#2#3{{
        \centering{\begin{minipage}{4.5in}\footnotesize\baselineskip=10pt
        \parindent=0pt #1\par 
        \parindent=15pt #2\par
        \parindent=15pt #3
        \end{minipage}}\par}} 

\renewenvironment{thebibliography}[1]
        {\frenchspacing
         \ninerm\baselineskip=11pt
         \begin{list}{\arabic{enumi}.}
        {\usecounter{enumi}\setlength{\parsep}{0pt}
         \setlength{\leftmargin 12.7pt}{\rightmargin 0pt} 
         \setlength{\itemsep}{0pt} \settowidth
        {\labelwidth}{#1.}\sloppy}}{\end{list}}

\newcounter{itemlistc}
\newcounter{romanlistc}
\newcounter{alphlistc}
\newcounter{arabiclistc}
\newenvironment{itemlist}
        {\setcounter{itemlistc}{0}
         \begin{list}{$\bullet$}
        {\usecounter{itemlistc}
         \setlength{\parsep}{0pt}
         \setlength{\itemsep}{0pt}}}{\end{list}}

\newcommand{\fcaption}[1]{
        \refstepcounter{figure}
        \setbox\@tempboxa = \hbox{\footnotesize Fig.~\thefigure. #1}
        \ifdim \wd\@tempboxa > 5in
           {\begin{center}
        \parbox{5in}{\footnotesize\smalllineskip Fig.~\thefigure. #1}
            \end{center}}
        \else
             {\begin{center}
             {\footnotesize Fig.~\thefigure. #1}
              \end{center}}
        \fi}

\newcommand{\tcaption}[1]{
        \refstepcounter{table}
        \setbox\@tempboxa = \hbox{\footnotesize Table~\thetable. #1}
        \ifdim \wd\@tempboxa > 5in
           {\begin{center}
        \parbox{5in}{\footnotesize\smalllineskip Table~\thetable. #1}
            \end{center}}
        \else
             {\begin{center}
             {\footnotesize Table~\thetable. #1}
              \end{center}}
        \fi}

\def\@citex[#1]#2{\if@filesw\immediate\write\@auxout
        {\string\citation{#2}}\fi
\def\@citea{}\@cite{\@for\@citeb:=#2\do
        {\@citea\def\@citea{,}\@ifundefined
        {b@\@citeb}{{\bf ?}\@warning
        {Citation `\@citeb' on page \thepage \space undefined}}
        {\csname b@\@citeb\endcsname}}}{#1}}

\newif\if@cghi
\def\cite{\@cghitrue\@ifnextchar [{\@tempswatrue
        \@citex}{\@tempswafalse\@citex[]}}
\def\citelow{\@cghifalse\@ifnextchar [{\@tempswatrue
        \@citex}{\@tempswafalse\@citex[]}}
\def\@cite#1#2{{$\null^{#1}$\if@tempswa\typeout
        {IJCGA warning: optional citation argument 
        ignored: `#2'} \fi}}

\def\pmb#1{\setbox0=\hbox{#1}
        \kern-.025em\copy0\kern-\wd0
        \kern.05em\copy0\kern-\wd0
        \kern-.025em\raise.0433em\box0}

\def\fnt#1#2{\footnotetext{\kern-.3em
        {$^{\mbox{\scriptsize #1}}$}{#2}}}

\def\thefootnote{\fnsymbol{footnote}}
\def\@makefnmark{\hbox to 0pt{$^{\@thefnmark}$\hss}}    
        
\def\ps@myheadings{%
    \let\@oddfoot\@empty\let\@evenfoot\@empty
    \def\@evenhead{\slshape\leftmark\hfil}
    \def\@oddhead{\hfil{\slshape\rightmark}}
    \let\@mkboth\@gobbletwo
    \let\sectionmark\@gobble
    \let\subsectionmark\@gobble
    }
\font\tenrm=cmr10
\font\tenit=cmti10 
\font\tenbf=cmbx10
\font\bfit=cmbxti10 at 10pt
\font\ninerm=cmr9

\font\eightrm=cmr8

\def\qed{\hbox{${\vcenter{\vbox{                        
   \hrule height 0.4pt\hbox{\vrule width 0.4pt height 6pt
   \kern5pt\vrule width 0.4pt}\hrule height 0.4pt}}}$}}

\renewcommand{\thefootnote}{\fnsymbol{footnote}}  

\begin{document}
\setlength{\textheight}{7.7truein}  

\thispagestyle{empty}

\markboth{\protect{\footnotesize\it Loops and Loop
    Clouds}}{\protect{\footnotesize\it Loops and Loop Clouds}}

\normalsize\textlineskip

\setcounter{page}{1}

\copyrightheading{}             

\vspace*{1cm}

\centerline{\bf LOOPS AND LOOP CLOUDS}
\vspace*{0.035truein}
\centerline{\bf -- A NUMERICAL APPROACH}
\centerline{\bf  TO THE WORLDLINE FORMALISM IN QED --\footnote{Talk
    given by H.~Gies at the {\em Fifth Workshop on Quantum Field Theory
      under the Influence of External Conditions}, Leipzig, Germany,
    September, 2001.}}
\vspace*{0.37truein}
\centerline{\footnotesize Holger Gies}
\baselineskip=12pt
\centerline{\footnotesize\it CERN, Theory Division}
\baselineskip=10pt
\centerline{\footnotesize\it CH-1211 Geneva 23,
  Switzerland}

\vspace*{10pt}
\centerline{\footnotesize Kurt Langfeld}
\baselineskip=12pt
\centerline{\footnotesize\it Institut f\"ur theoretische Physik,
  Universit\"at T\"ubingen}
\baselineskip=10pt
\centerline{\footnotesize\it D-72076 T\"ubingen, Germany}

\vspace*{0.21truein}
\abstracts{A numerical technique for calculating effective actions of
  electromagnetic backgrounds is proposed, which is based on the
  string-inspired worldline formalism. As examples, we consider scalar 
  electrodynamics in three and four dimensions to one-loop
  order. Beyond the constant-magnetic-field case, we analyze
  a step-function-like magnetic field exhibiting a nonlocal and
  nonperturbative phenomenon: ``magnetic-field diffusion''. Finally,
  generalizations to fermionic loops and systems at finite temperature
  are discussed.}{}{}


\vspace*{1pt}\textlineskip      
\section{Introduction}  
\vspace*{-0.5pt}
\noindent
The computation of effective energies and effective actions of a
quantum system is a general and frequently occurring problem in quantum
field theory. In a given classical background, virtual quantum
processes such as vacuum polarization modify the properties of the
vacuum and are responsible for new nonlinear and nonlocal phenomena.
These phenomena can often be described by an effective action that
takes the virtual quantum processes into account. A prominent example
is given by the Heisenberg-Euler
action\cite{Heisenberg:1935qt,Schwinger:1951nm}, describing the nonlinear
corrections to the Maxwell action which are induced by virtual loops
of electrons and positrons.

The Heisenberg-Euler action is the one-loop QED effective action,
\begin{equation}
\Gamma_{\text{eff}}^{\text{spinor}}=\ln \det (-\I\fss{\partial}
+\fss{A} -\I m), \quad \Gamma_{\text{eff}}^{\text{scalar}}=-\ln \det
( -(\partial +\I  A)^2+m^2), \label{1}
\end{equation}
(here in Euclidean formulation), evaluated for a constant
electromagnetic background field, and thereby represents a valid
approximation of low-energy QED for all backgrounds that vary
slowly with respect to the Compton wavelength $1/m$.

Technical difficulties increase markedly if one wants to go beyond
this approximation. The standard means is the derivative
expansion\cite{Gusynin:1999bt}; its application is however limited to
backgrounds whose spacetime variation does not exceed the scale of
the electron mass or the scale of the field strength. Moreover, the
enormous proliferation of terms restricts the actual computations to
low orders. Only for very special (and very rare) field configurations
can the derivative expansion be summed up\cite{Cangemi:1995ee}.
Numerical methods developed so far aim at the integration of the
corresponding differential equation of the operators in Eq.\re{1};
therefore, they have been applied only to highly symmetric background
fields up to now\cite{Bordag:1998tg}. Of course, also the brute force
method of spacetime discretization (lattice) can be employed, which
has its own shortcomings (finite-size problems, fermion doubling,
problems with many scales).

In this note, we advocate a recently developed numerical
technique\cite{Gies:2001zp} which is based on the string-inspired
worldline formalism\cite{berkos} applied to QED with background
fields\cite{Schmidt:1993rk}. The idea of the approach consists of
rewriting the functional determinants contained in Eq.\re{1} in terms
of 1-dim.  path integrals, which finally can be evaluated with
Monte-Carlo techniques. In contrast to the techniques mentioned above,
our worldline numerics makes no reference to the particular properties
of the background field and is therefore applicable to a wide class of
configurations.  In the following, we describe the approach in more
detail and list a number of examples of a basic nature.

\setcounter{footnote}{0}
\renewcommand{\thefootnote}{\alph{footnote}}

\section{Basics of worldline numerics}
\label{sec2}
\noindent
In the following, we first concentrate on scalar QED, starting with
the unrenormalized, but regularized Euclidean one-loop effective
action in $D$ dimensions in worldline
representation\cite{Schubert:2001he}, corresponding to the second
equation of\re{1},
\begin{equation} 
\Geff^{\text{scalar}}[A]
=\int\limits_{1/\Lambda^2}^\infty \frac{dT}{T}\, \E^{-m^2
  T}\, {\cal N}   
\int\limits_{x(T)=x(0)} {\cal D}x(\tau)\, \E^{-\int\limits_0^T d\tau 
  \left( \frac{\dot{x}^2}{4} +\I e\,\dot{x}\cdot
    A(x(\tau))\right)}. \label{1a} 
\end{equation} 
Here we encounter a path integral over closed loops in
spacetime. Note that there are no other constraints to the loops
except differentiability and closure; in particular, they can be
arbitrarily self-intersecting and knotty. Introducing the Wilson loop
\begin{equation} 
W[A(x)]=\E^{-\I e\int\limits_0^T d\tau\, 
  \dot{x}(\tau) \cdot A(x(\tau))} \equiv \E^{-\I e\oint dx\cdot A(x)},
  \label{4}   
\end{equation}
Eq.\re{1a} can be rewritten in a compact form:
\begin{equation} 
\Geff^{\text{scalar}}[A]=\frac{1}{(4\pi)^{D/2}} \int d^Dx_0
\int\limits_{1/\Lambda^2}^\infty  
\frac{dT}{T^{(D/2)+1}}\, \E^{-m^2 T} \langle W[A]\rangle_x. \label{3} 
\end{equation}
Here we also have split off the integral over the loop centers of
mass, $x_0^\mu:=(1/T)\int_0^Td\tau\,x^\mu(\tau)$, and
$\langle(\dots)\rangle_x$ denotes the expectation value of (\dots)
evaluated over an ensemble of $x$ loops; the loops are centered upon a
common average position $x_0$ (``center of mass'') and are distributed
according to the Gaussian weight $\exp[-\int_0^T d\tau\,
\frac{\dot{x}^2}{4}]$.

Crucial for the numerical realization is the observation that
substituting $\tau=:T t$ and introducing {\em unit loops} $y$, 
\begin{equation} 
y(t):=\frac{1}{\sqrt{T}}\, x(T t), \quad t\in [0,1], \label{5} 
\end{equation} 
renders the the Gaussian weight independent of the propertime:
$\int_0^T\!d\tau\, {\dot{x}^2(\tau)}/{4} = \int_0^1\!dt\,
{\dot{y}^2(t)}/{4}$ (the dot ``$\dot{\quad}$'' denotes always a
derivative with respect to the argument). Now, the expectation value
of $W[A]$ can be evaluated over the unit-loop ensemble $y$,
\begin{equation} 
\langle W[A(x)]\rangle_x \equiv \langle 
W[\sqrt{T}A(x_0+\sqrt{T}y)]\rangle_y,\label{7} 
\end{equation} 
where the exterior $T$-propertime dependence occurs only as a scaling
factor of the gauge field and the unit loops in its argument. In other
words, while approximating the loop path integral by a finite ensemble
of loops, it suffices to have one single unit-loop ensemble at our
disposal; we do not have to generate a new loop ensemble whenever we
go over to a new value of $T$.

\subsection{Renormalization}
\noindent
The effective action is renormalized by adding counterterms to $\Geff$
in order to absorb the strongly cutoff-dependent parts. The
counterterms are determined by a set of physical constraints such as
the vanishing of $\Geff$ for a vanishing background and the value of
the gauge coupling in soft processes at a certain scale. To be
specific, we use standard Coleman-Weinberg renormalization conditions
for the complete action $\Gamma$, containing the bare Maxwell action
and the one-loop contribution of Eq.\re{1},
\begin{equation}
\Gamma[F=0]=0,\quad \frac{\delta \Gamma}{\delta
(F^2)}\bigg|_{F^2/e_{\text{R}}^2=\mu^2/2} =\frac{1}{4\,
e_{\text{R}}^2(\mu)}, \label{renormcond}
\end{equation}
where we identify $\mu$ with the scale of soft photons measuring the
Thomson cross section, i.e., $\mu/m\to 0$. In the examples of
Sect.~3, we will only encounter the field strength $F$
which is renormalization group invariant, since it is scaled by the
coupling. The substitution $F\to e_{\text{R}}F_{\text{R}}$ reexpresses
the later results in terms of physical coupling and field strength,
where $e_{\text{R}}$ and $F_{\text{R}}$ are the renormalized
quantities at the scale $\mu$ mentioned above. 

From a technical viewpoint, the strongly $\Lambda$-dependent terms can
be isolated using a heat-kernel expansion of the Wilson loop (which
can be performed for arbitrary backgrounds), e.g.,
\begin{equation}  
\biggl\langle W[\sqrt{T}A(x_0 +\sqrt{T}y)] \biggr\rangle_y \; = \;  
1 \; - \; \frac{1}{12} T^2 \, F_{\mu \nu }[A] (x_0) \, F_{\mu \nu }[A]
(x_0) \; + \; {\cal O} (T^4) \; .
\label{10}  
\end{equation}
The counterterms required are in one-to-one correspondence to the
terms of the small-$T$ expansion, encoding the short-distance
physics; the number of necessary counterterms depends on the dimension
of spacetime. While this renormalization program is well under control
analytically, the numerical renormalization is complicated by a
further problem: evaluating 
$\langle W\rangle$ with the aid of the loop ensemble does not produce 
the small-$T$ behavior of Eq.\re{10} {\em exactly}, but, of course, 
only within the numerical accuracy. Unfortunately, even the smallest 
deviation from Eq.\re{10} will lead to huge errors of the effective
action, because it induces an artificial singular behavior of the
propertime integrand. 

Our solution to this problem is to fit the numerical result for
$\langle W\rangle$ to a polynomial in $T$ in the vicinity of $T=0$,
employing Eq.\re{10} as a constraint for the first coefficients. In
other words, we insert the analytical information about the short
distance behavior of $\langle W\rangle$ for small $T$ explicitly. This
fit not only isolates the strongly cutoff-dependent parts which
subsequently are subject to the standard renormalization procedure,
but also facilitates a more precise estimate of the error bars.
Finally, employing this fitting procedure only close to $T=0$, the
infrared behavior ($T\to \infty$) of the integrand remains untouched,
and our approach is immediately applicable, also in the case $m=0$.

\subsection{Numerical simulation }  
\noindent
The numerical computation of the effective action can be summarized by
the following recipe:
\begin{itemlist} 
\item[(1)] generate a unit loop ensemble distributed according to the 
  weight $\exp[ -\int_0^1dt \dot{y}^2/4]$, e.g., employing the 
  technique of normal (Gaussian) deviates; 
\item[(2)] compute the integrand for arbitrary values of $T$ (and 
  $x_0$); this involves the evaluation of the Wilson loop expectation 
  value for a given background;  
\item[(3)] perform the renormalization procedure; 
\item[(4)] integrate over the propertime $T$ in order to obtain the 
Lagrangian, and also over $x_0$ for the action. 
\end{itemlist} 
There are two sources of error which are introduced by reducing the
degrees of freedom from an infinite to a finite amount: first, the
loop path integral has to be approximated by a finite number of loops;
second, the propertime $t$ of each loop has to be discretized.
Contrary to this, the spacetime does not require discretization, i.e.,
the loop ensemble is generated in the continuum. 

Approximating $\langle W \rangle $ of Eq.~(\ref{7}) by an average over
a finite number $N_L$ of loops, the standard deviation provides an
estimate of the statistical error. Approximating the loops by the
finite number $n_l$ of space points results in a systematic error that
can be estimated by repeating the calculation for several values
$n_l$. The number $n_l$ should be chosen large enough to reduce this
systematic error to well below the statistical one.  It will turn out
that the choice $N_L=1000$ and $n_l=100$ for $D=3$ ($n_l=200$ for
$D=4$) yields results, for the applications below, which are accurate
at the per cent level. 

\subsection{Loop Clouds}
\begin{figure}
\begin{picture}(120,30)
\put(0,30){(a)}
\put(72,30){(b)}
\put(26,14){\epsfig{figure=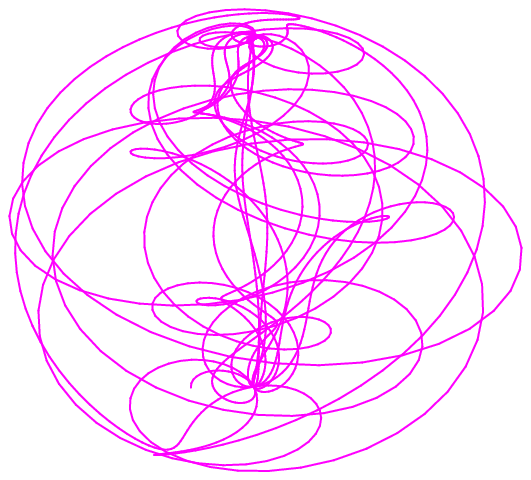,width=1.5cm}}
\put(77,0){\epsfig{figure=cloudy.eps,width=4cm}} 
\put(5,0){\epsfig{figure=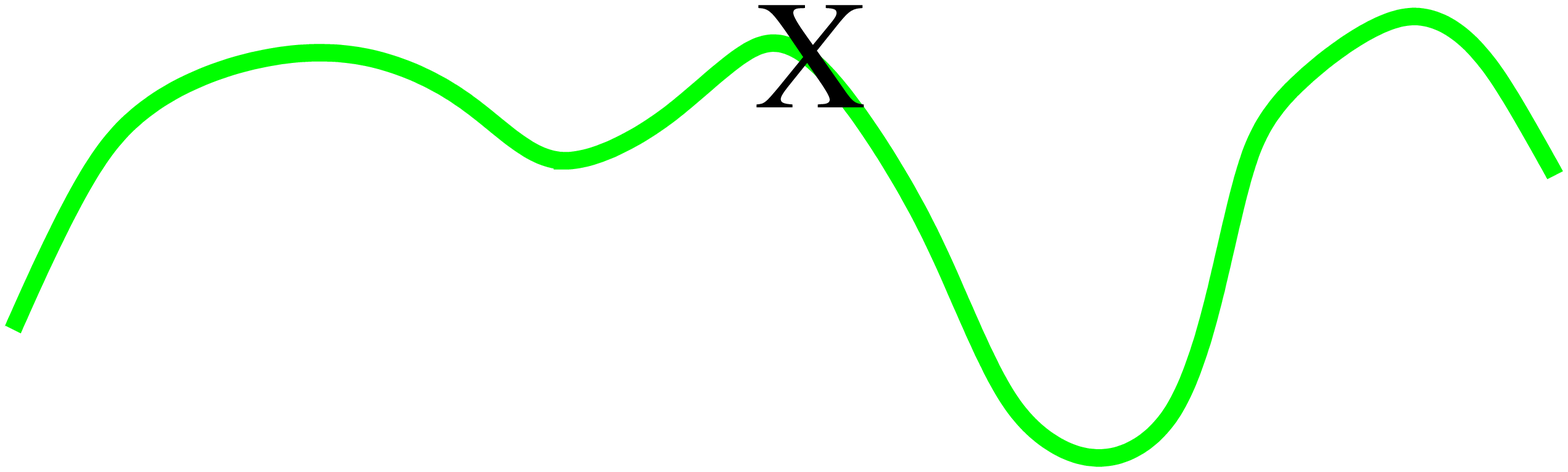,width=5.5cm,height=2.5cm}}
\put(70,0){\epsfig{figure=Fcontour.eps,width=5.5cm,height=2.5cm}}
\put(25,0){\Large {UV}}
\put(90,0){\Large {IR}}
\end{picture}
\caption{(a/b) small/large-propertime contribution to the
  effective-action density, picking up small/large-scale information
  of the background field (UV/IR).}
\label{zoom}
\end{figure}

\noindent
Beyond any numerical efficiency, worldline numerics also offers an
intuitive approach to effective Lagrangians or functional
determinants: Formula\re{3} together with\re{7} provides for a
descriptive interpretation of the quantum processes. For a given
background, the effective-action density receives contributions from
all values of the propertime. For instance, for small propertimes,
the size of the ensemble of unit loops is also small by virtue of the
propertime scaling in Eq.\re{7}, $\sqrt{T}y$. Hence, this {\em loop
  cloud} picks up small-scale information about the background
field. By contrast, for large propertimes, the loop cloud becomes
bloated and receives information about the behavior of the background
field over large scales. This zooming in or out of the quantum vacuum
is illustrated in Fig.~\ref{zoom}. 

Analyzing vacuum polarization in a certain background with this
concept of loop clouds thereby visualizes the nonlinearities and
nonlocalities of the quantum effective action in a vivid way.

\section{Examples}
\label{examples}

\vspace{-.3cm}

\subsection{Constant magnetic background field}
\noindent
Let us first investigate the efficiency of our numerical loop approach
to the scalar functional determinant for a constant
magnetic background field $B$. For this case, the Wilson loop
expectation value is exactly known and independent of $D$:
\begin{equation}
\langle W[A]\rangle=\frac{B T}{\sinh B T}, \quad \text{for}\quad
B=\text{const.}\label{const}
\end{equation}
Integration over the propertime leads us to the effective-action
density, which we plotted in Fig.\re{fig:1} for $D=3$ and $D=4$. The
agreement between the exact and the numerical results is satisfactory;
the exact results lie well within the error bars, which contain the
statistical errors of the Monte-Carlo calculation as well as those of
the fitting procedure near $T=0$ (the latter actually improve the
error estimate, since our exact knowledge about the $T=0$ behavior has
been inserted).
\begin{figure}[h] 
\begin{picture}(120,41) 
\put(2,-4){\epsfig{figure=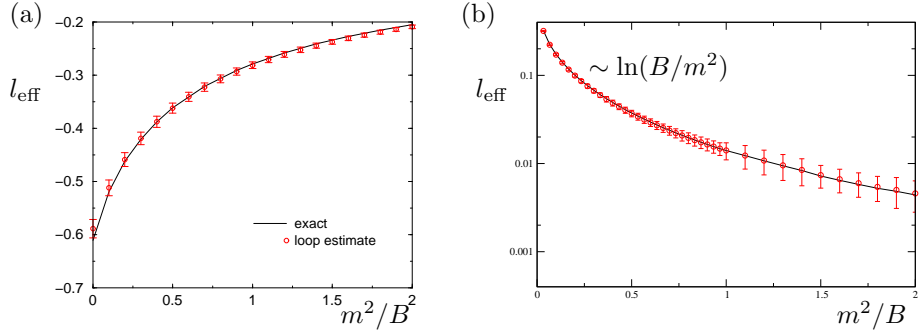,width=5.2cm}}
\put(0,38){(a)}
\put(70,-1){\epsfig{figure=Geff4.eps,width=3.7cm}}
\put(60,38){(b)} 
\put(0,28){$l_{\text{eff}}$}
\put(44,-2){$m^2/B$}
\put(62,28){$l_{\text{eff}}$}
\put(110,-2){$m^2/B$}
\put(77,31){$\sim\ln (B/m^2)$}
\end{picture}
\caption{One-loop effective-action density $l_{\text{eff}}$ in units
  of $(B/(4\pi))^{D/2}$ in $D=3$ (a) and $D=4$ (b) versus $m^2/B$ for
  the case of a constant magnetic background field. The analytically
  known exact results (solid lines) are compared with the numerical
  findings (circles with error bars).}
\label{fig:1}  
\end{figure}  
Our approach is able to cover a wide range of parameter values: in
$D=3$, even the massless limit can be taken without problems. In
$D=4$, the typical logarithmic increase in the strong-field
limit with a prefactor proportional to the $\beta$ function is
visible. 
 
\subsection{Step-like magnetic field: quantum diffusion}
\noindent
As a specific example for a field configuration that cannot be treated
in a derivative expansion but inherently involves nonlocal aspects, we
consider a time-like constant background field $B$, resembling
a step function in space,
\begin{equation}  
B(x,y) \; = \; - \theta (x) \, B   \; ,  
\quad 
\vec{A}(x,y) \; = \; \theta (x) \; \frac{1}{2} \, (y,-x) \; B \; ,  
\label{eq:d1}  
\end{equation} 
working for simplicity in $D=3$. In our approach, discontinuities do
not induce (artificial) singularities, but are smoothly controlled by
the properties of the loop ensemble. This loop cloud has finite
extension and slowly varying density; while running with its center of
mass towards and across the step, that part of the volume of the loop
cloud which ``feels'' the magnetic field increases smoothly. 

\begin{figure}[h] 
\begin{picture}(120,41) 
\put(0,-2){\epsfig{figure=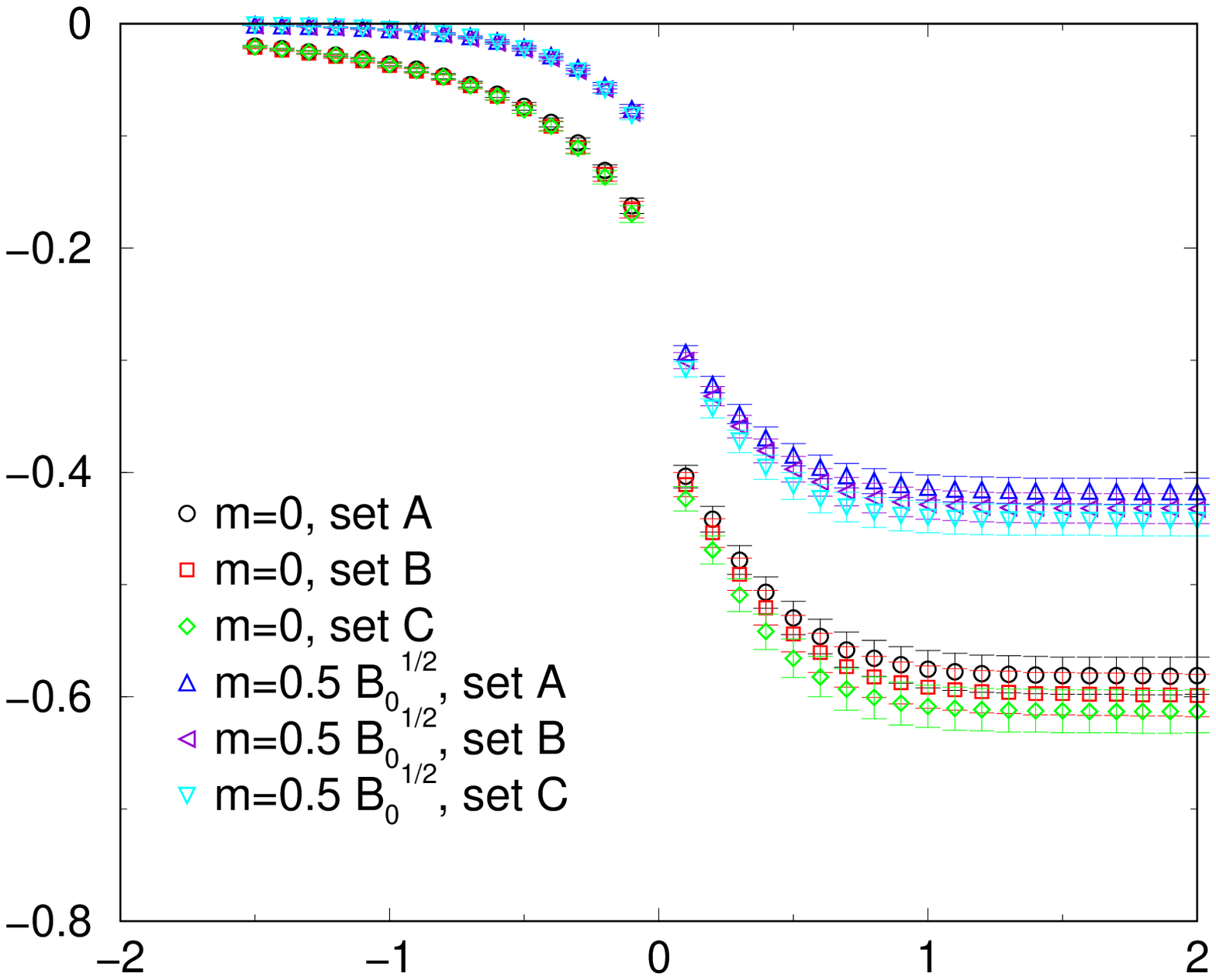,width=5.2cm}}
\put(0,38){(a)}
\put(65,-4){\epsfig{figure=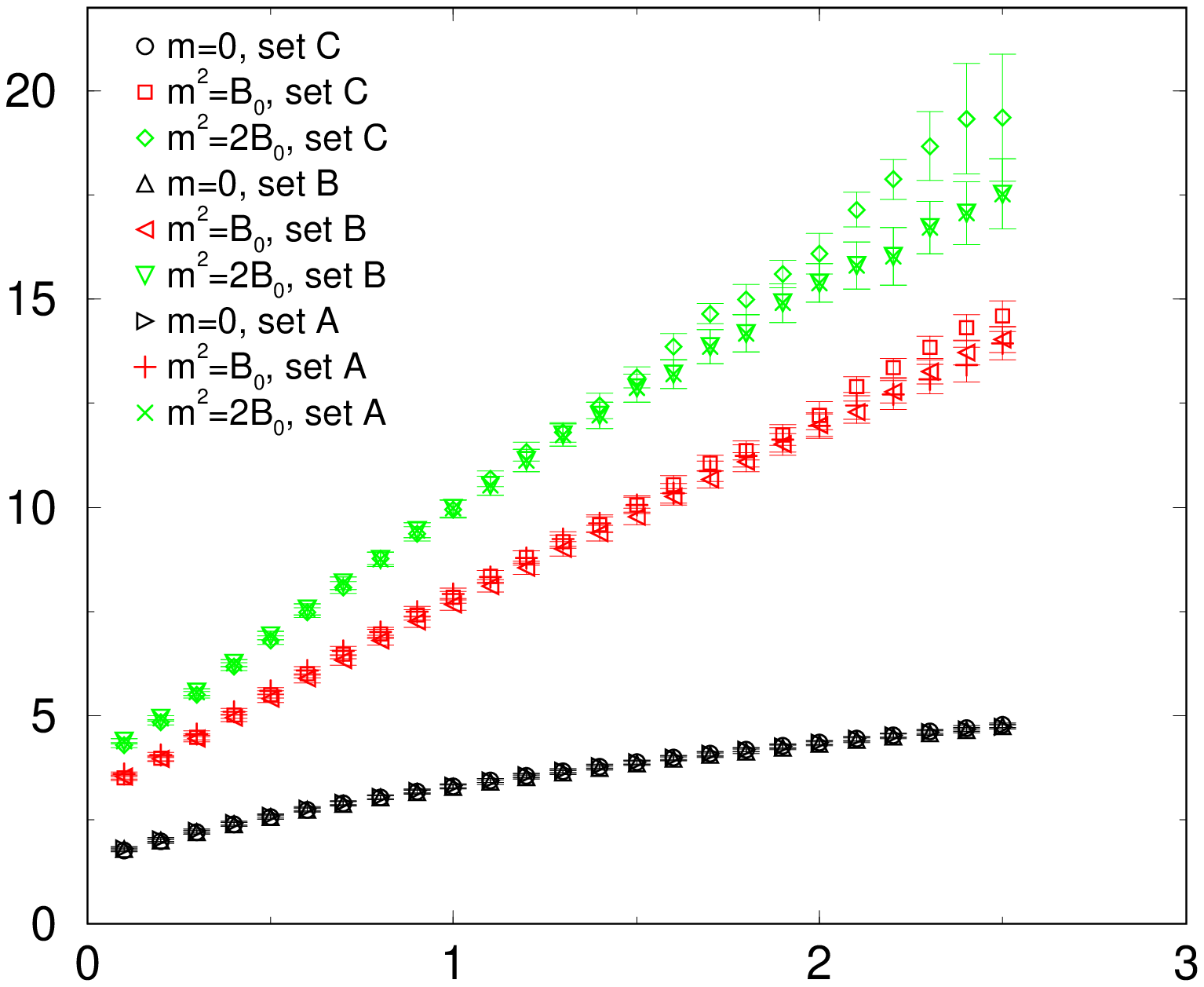,width=5.2cm}}
\put(60,38){(b)}
\put(0,28){$l_{\text{eff}}$}
\put(58,31){$\ln (-l_{\text{eff}})$}
\put(52,-2){$x\sqrt{B}$}
\put(113,-2){$-x\sqrt{B}$} 
\end{picture}
\caption{(a) Diffusion profile of the effective-action density in the
  vicinity of the magnetic step ($x=0$) in units of $(B/(4\pi))^{3/2}$
  for various mass-to-field-strength ratios. (b) A logarithmic plot
  reveals the exponential nature of the diffusion depth. For a check
  of the propertime-continuum limit, three different loop ensembles
  have been used: $n_l=100$ for set A, $n_l=75$ for set B, $n_l=50$
  for set C.}
\label{fig:2} 
\end{figure}

In Fig.~\ref{fig:2}, we plot our numerical result for the
effective-action density across the step (for details,
see\cite{Gies:2001zp}). As expected, the effective-action density is
nonzero even in the region $x<0$ where the background field
$B(\vec{x})$ vanishes.
In order to characterize this {\em quantum diffusion} of the magnetic
field further, we fit the effective-action density $\Leff$ in the
zero-field region using the ansatz\footnote{The spacetime coordinates
  occuring here are, of course, identical to the center-of-mass
  coordinates $x_0$ of the worldlines (cf.~Sect.~2); we dropped the
  subscript for simplicity.}
\begin{equation}
l_{\text{eff}}=\frac{(4\pi)^{3/2}\,\Leff}{B^{3/2}}\sim \exp \bigl(
-\beta_1\, m\, |x| -\beta_2\, \sqrt{B}\, |x|\bigr). \label{fit}
\end{equation}
In fact, the numbers $\beta_1\simeq3.255$ and $\beta_2\simeq0.7627$
fit the data with a high accuracy. The occurrence of two different
diffusion lengths $\sim 1/(\beta_1 m)$ and $\sim 1/(\beta_2 \sqrt{B})$
may come as a surprise, since one might have expected that the only
scale in the field-free region is given by the mass $m$. However,
since the loop cloud is an extended object, it will ``feel'' the
magnetic field even at a large distance; hence, the magnetic field is
also a valid scale in the field-free region. This is precisely a
consequence of the inherent nonlocality of the quantum processes.
Moreover, the diffusion phenomenon appears to be also nonperturbative
in the same sense as the Schwinger mechanism of pair
production\cite{Schwinger:1951nm}. This is suggested by the functional
form Eq.\re{fit}, which cannot be expanded in terms of the coupling
constant, being rescaled in the field (only an expansion in terms of
the square root of the coupling constant is possible).

\subsection{Constant magnetic field at finite temperature}
\noindent
The presence of a heat bath can be taken into account within the
worldline framework by using the imaginary-time formalism: the
Euclidean time direction is compactified to a circle with
circumference
$\beta=1/$temperature\cite{Shovkovy:1998xw,Gies:1998vt}.\footnote{The
  propertime $T$ should not be confused with the temperature, being
  characterized by its inverse $\beta$ in the following.}\quad This
changes the boundary conditions of the worldline; the loops do not
have to close trivially in spacetime, but can now wind around the
spacetime cylinder $n$ times:
\begin{equation}
\int\limits_{x(T)=x(0)}{\cal D}x\quad \to\quad
\sum_{n=-\infty}^\infty\,\,\int\limits_{\vec{x}(T)=\vec{x}(0),
  x_0(T)=x_0(0)+n\beta}{\cal D}x.\label{wind}
\end{equation}
Choosing a gauge where the time-like component of the gauge field is
time independent, a Poisson resummation of the $n$-sum disentangles
the winding around the cylinder from the trivially closed loops for
any background, and the remaining path integral runs over nonwinding
loops only. The numerical problem thereby reduces to the
zero-temperature case with an additional summation over Matsubara
modes, which can be performed to a high numerical accuracy.

\begin{figure}[h]
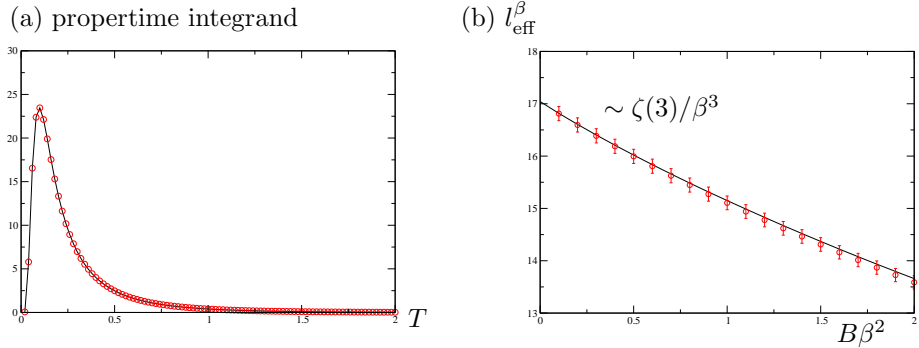
 
\begin{picture}(120,43) 
\put(1,0){\epsfig{figure=integrandFinT.eps,width=3.7cm}}
\put(0,40){(a) propertime integrand}
\put(70,0){\epsfig{figure=GeffFin.eps,width=3.7cm}}
\put(60,40){(b) $l_{\text{eff}}^\beta$}
\put(53,0){$T$}
\put(79,28){$\sim\zeta(3)/\beta^3$}
\put(110,-2){$B \beta^2$}
\end{picture}
\caption{(a) Propertime integrand of the finite-temperature
  contribution to the effective-action density in $D=3$ scalar QED
  ($m=1/\beta$, $m^2=B$). (b) Effective-action density contribution at
  finite temperature in units of $2/(4\pi\beta^2)^{3/2}$ versus
  $B\beta^2$ in the massless limit.}
\label{fintfig}
\end{figure}

Confining ourselves to the constant-magnetic field case, the exact
result for the finite-temperature contribution to the Wilson loop EV
reads \cite{Gies:1998vt,Dittrich:2000zu}
\begin{equation}
\langle W[A]\rangle^\beta_x=2\sum_{n=1}^\infty\frac{T B}{\sinh TB}\,
\E^{-\frac{n^2\beta^2}{4T}}. \label{fintWL}
\end{equation}
Concentrating on $D=3$ scalar QED, we compare the exact results with
the numerical estimates in Fig.~\ref{fintfig}. The propertime
integrand (plotted for $m=1/\beta$, $m^2=B$) as well as the
effective-action density (plotted for $m=0$) can satisfactorily be
reproduced by the numerical approach. The drop-off of the integrand
for small propertimes in Fig.~\ref{fintfig}(a) can be understood in
terms of the loop clouds: for small $T$ the loop cloud is simply too
small to ``see'' the compactness of the Euclidean time direction.
Furthermore, the Stefan-Boltzmann law for the free energy ($=-$
effective-action density) $\sim \zeta(3)/\beta^3$ is rediscovered in
the $B=0$ limit.

\subsection{Fermions in the loop}
\noindent
The worldline formalism allows for an elegant inclusion of fermions by
representing the Dirac algebraic elements by path integrals over
closed loops in Grassmann space. Unfortunately, this cannot be
translated into a numerical algorithm, since the kinetic action for
the Grassmann loops is not positive definite; this prohibits a
Monte-Carlo algorithm based on importance sampling. Instead, we choose
to maintain the Dirac algebraic formulation, which leads to the
following representation of the spinor part of Eq.\re{1}:
\begin{equation}
\Geff^{\text{spinor}}=-\frac12\frac{1}{(4\pi)^{D/2}}
\int\limits_{1/\Lambda^2}^\infty  \frac{dT}{T}\, \E^{-m^2
  T}\,\big\langle W_{\text{spin}}[A]\big\rangle_x, \label{spin1}
\end{equation}
where the ``spinorial'' Wilson loop is given by
\begin{equation}
 W_{\text{spin}}[A]=W[A]\times\,\Tr\, P_T\, \exp\left(\frac{\I e}{2}
 \int_0^T d\tau\, \sigma_{\mu\nu} F^{\mu\nu}\right). \label{spin2}
\end{equation}
Obviously, the spinorial Wilson loop factorizes into the usual one and
a trace of a path-ordered exponential of the spin-field coupling.
Although this path ordering is difficult to handle in analytical
computations, our numerical algorithm can easily deal with; this
is because the path(=loop) is discretized and each point is visited in
a path-ordered manner anyway. As a further simplification, the
exponential of the spin-field coupling can be projected onto an
orthogonal basis of the Dirac algebra (see, e.g.,
\cite{Dittrich:2000zu}).

\begin{figure}[h]
\begin{picture}(120,37)
\put(20,-3){
\epsfig{figure=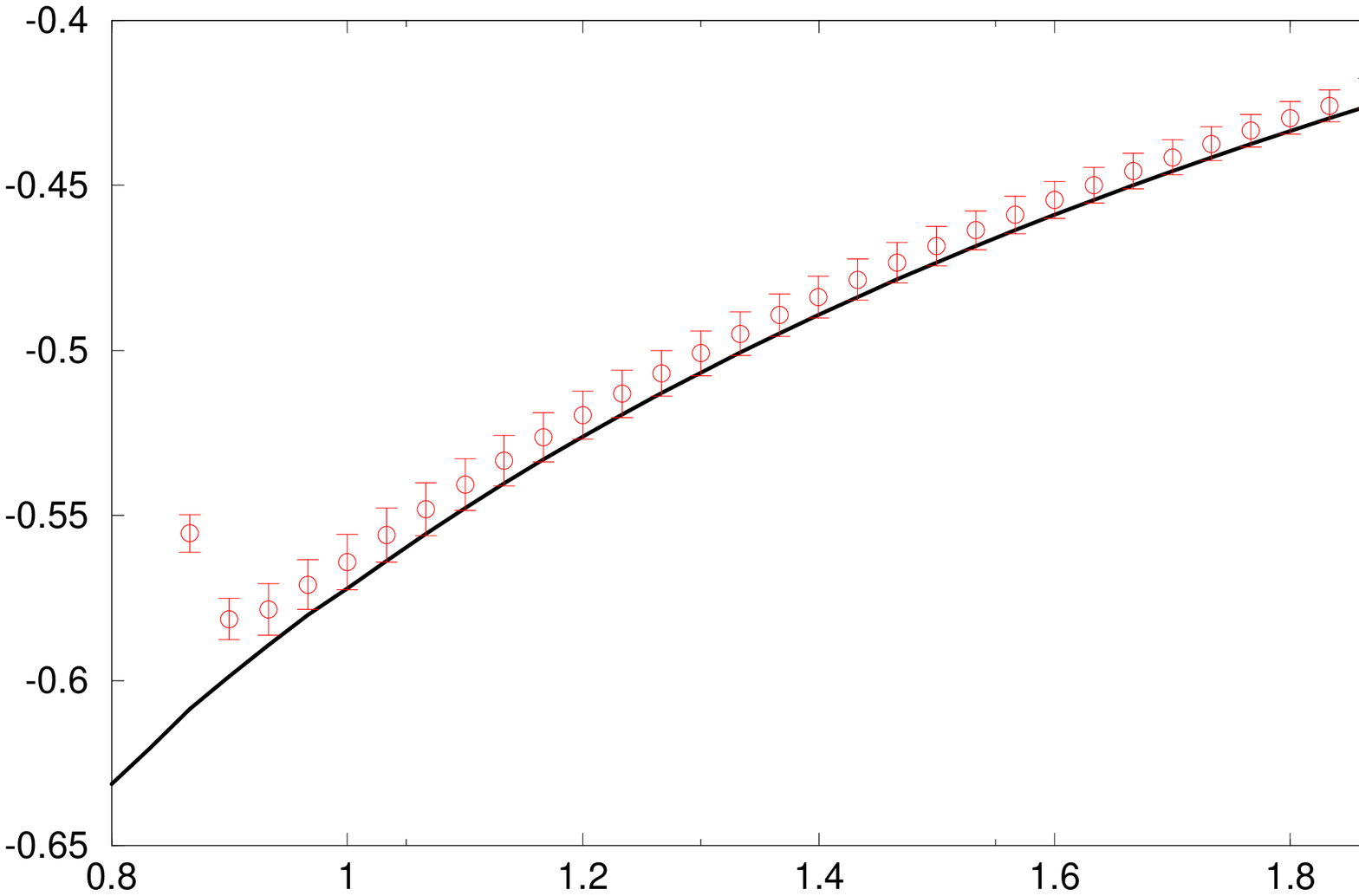,width=6cm,height=4cm}}
\put(15,25){$l_{\text{eff}}^{\text{spinor}}$}
\put(90,-3){$m^2/B$}
\end{picture}
\caption{Effective-action density of spinor QED in $D=3$ for the 
constant-magnetic-field case (units as in Fig.~\ref{fig:1}).}
\label{figferm}
\end{figure}

Concentrating again on the constant-magnetic-field case, the
analytically known result for the spinorial Wilson loop is
\begin{equation}
\big \langle W_{\text{spin}}[A]\big\rangle=BT\coth BT\equiv \frac{
 BT}{\sinh BT}\,\cosh BT,\label{spin3}
\end{equation}
where the cosh term arises from the trace of the path-ordered
exponential in Eq.\re{spin2}. Comparing the exact result with the
numerical one in $D=3$ as depicted in Fig.\ref{figferm}, the agreement
in the parameter range $B\lesssim m^2$ is again satisfactory. Only in
the strong-field limit does the algorithm become unstable owing to
technical reasons: here the exponential increases of the cosh and the
sinh function do not cancel numerically as precisely as they should.
This technical problem can be solved by first matching the exponential
increases of the denominator and the numerator to each other and then
performing the division. Whether this matching can be numerically
implemented in a background-field independent way has still to be
investigated. Beyond this, there are no problems of any fundamental
kind limiting the application of the present approach to fermion 
determinants, in contrast to no-go theorems faced by lattice formulations.  

\newpage

\section{Conclusion}
\noindent                       %
Our results of the numerical worldline approach to QED effective
actions are encouraging: the successful description of the background
conditions selected here indicates that the method can be applied to a
very general class of backgrounds without the necessity of assuming
slow variation, smoothness, or symmetries. In particular, it should be
stressed that the algorithm makes no reference to the properties of
the background field.

Beyond being an efficient method of computation, our approach also
offers a vivid picture of the quantum world: consider a spacetime
point $x$ at a propertime $T$; then, the loop cloud is centered upon
this point $x$ with Gaussian ``density'' and ``spread''. Increasing or
lowering the propertime $T$ corresponds to bloating or scaling down
the loop cloud or, alternatively, zooming out of or into the
microscopic world. The effective-action density at each point $x$
finally receives contributions from every point of the loop cloud
according to its Gaussian weight and averaged over the propertime.
This gives rise to the inherent nonlocality and nonlinearity of the
effective action, because every point $x$ is influenced by the field
of any other point in spacetime experienced by the loop cloud.

We would like to conclude with some further remarks and caveats. Of
course, the Monte-Carlo procedure requires that the loops are
distributed in Euclidean spacetime in order to guarantee the
positivity of the action.  This can lead to difficulties (also shared
by many other approaches) when the method is applied to the
computation of Green´s functions with Minkowskian momenta. However,
this does not pose a problem for the inclusion of Minkowskian electric
fields which can simply be taken into account by an imaginary
Euclidean field strength. For instance, pair production in
inhomogeneous electric fields can be studied in this way.  Finally,
the analogy between worldline path integrals and Brownian motion
implies that the loops in our loop cloud should have Hausdorff
dimension 2; on a computer, this can, of course, never be realized.
However, we expect that this may lead to sizeable errors only in the
extreme case of a massless or very light particle in the loop {\em
  and} a background field with fluctuations on all scales; only then
are the large-propertime contributions not suppressed, and the coarse
structure of the large loop clouds are insensitive to small-scale
fluctuations. This failure will manifest itself in the practical
impossibility of performing the propertime-continuum limit. On the
other hand, if the continuum limit can be taken, the loops will be
sufficiently close to Hausdorff dimension 2.

\nonumsection{Acknowledgements}
\noindent
H.G.~would like to thank M.~Bordag for organizing this productive
workshop and acknowledges discussions with W.~Dittrich, G.V.~Dunne,
R.L.~Jaffe, G.-L.~Lin, C.~Schubert, M.~Reuter, and H.~Weigel. H.G.~is
grateful for the warm hospitality of the IFAE at Barcelona Autonoma
U., where this manuscript was completed. 
This work was supported by the
DFG under contract Gi 328/1-1.

\nonumsection{References}

\end{document}